\documentclass[prl,twocolumn,showpacs,preprintnumbers,amsmath,amssymb]{revtex4}
\usepackage{graphicx}
\usepackage{color}
\usepackage{epsfig}
\usepackage{latexsym}
\usepackage{bm}

\begin{document}

\renewcommand{\ni}{{\noindent}}
\newcommand{\dprime}{{\prime\prime}}
\newcommand{\be}{\begin{equation}}
\newcommand{\ee}{\end{equation}}
\newcommand{\bea}{\begin{eqnarray}} 
\newcommand{\eea}{\end{eqnarray}}
\newcommand{\nn}{\nonumber} 
\newcommand{\bk}{{\bf k}}
\newcommand{\bQ}{{\bf Q}}
\newcommand{\q}{{\bf q}}
\newcommand{\s}{{\bf s}}
\newcommand{\bN}{{\bf \nabla}}
\newcommand{\bA}{{\bf A}}
\newcommand{\bE}{{\bf E}}
\newcommand{\bj}{{\bf j}}
\newcommand{\bJ}{{\bf J}}
\newcommand{\bs}{{\bf v}_s}
\newcommand{\bn}{{\bf v}_n}
\newcommand{\bv}{{\bf v}} 
\newcommand{\la}{\langle}
\newcommand{\ra}{\rangle} 
\newcommand{\dg}{\dagger}
\newcommand{\br}{{\bf{r}}} 
\newcommand{\brp}{{\bf{r}^\prime}} 
\newcommand{\bq}{{\bf{q}}}
\newcommand{\hx}{\hat{\bf x}} 
\newcommand{\hy}{\hat{\bf y}}
\newcommand{\bS}{{\bf S}} 
\newcommand{\cU}{{\cal U}}
\newcommand{\cD}{{\cal D}} 
\newcommand{\bR}{{\bf R}}
\newcommand{\pll}{\parallel} 
\newcommand{\sumr}{\sum_{\vr}} 
\newcommand{\cP}{{\cal P}} 
\newcommand{\cQ}{{\cal Q}} 
\newcommand{\cS}{{\cal S}}
\newcommand{\ua}{\uparrow} 
\newcommand{\da}{\downarrow}
\newcommand{\red}{\textcolor {red}}

\title{Ion transport through confined ion channels in the presence of immobile charges}

\author{Punyabrata Pradhan, Yariv Kafri, and Dov Levine}

\affiliation{ Physics Department, Technion - Israel Institute of Technology, Haifa, Israel}

\begin{abstract}

\noindent
We study charge transport in an ionic solution in a confined nanoscale geometry in the presence of an externally applied electric field and immobile background charges. For a range of parameters, the ion current shows non-monotonic behavior as a function of the external ion concentration. For small applied electric field, the ion transport can be understood from simple analytic arguments, which are supported by Monte Carlo simulation. The results qualitatively explain  measurements of ion current seen in a recent experiment on ion transport through a DNA-threaded nanopore (D. J. Bonthuis {\it et. al.}, Phys. Rev. Lett, {\bf 97}, 128104 (2006)).  

\end{abstract}


\maketitle


Because of its central role in maintaining the homeostasis of cells, ion transport through channels across cell membranes is of great importance \cite{Hille, Roux, Doyle, Meller}. 
In a system with free ions, such as an aqueous solution, one might expect the ion current $I$ to increase with increasing external ion concentration $c$ when a constant electric field is applied.  Surprisingly, in the presence of immobile charges fixed in the channel, the opposite may occur, with ion current decreasing with increasing $c$. For example, in the case of water-filled biological channels with strong ion binding sites, the ion conductance has been observed to reach a maximum and then decrease (or saturate) as $c$ increases \cite{Hladky}; similar behavior is observed in DNA-threaded nanopores connecting two reservoirs \cite{Bonthuis}. 

An ion channel may be thought of as a thin hollow tube of length $L$ where ions can enter or leave only through pores at the two ends.  Because of the large difference in the dielectric constants of water ($\kappa_{w}\approx 80$) and the membrane containing the channel ($\kappa_{m} \approx 2$), introducing an uncompensated ion into the channel requires overcoming an energy barrier due to the charge's self energy $U_{S}$ \cite{Parsegian}. The reason for this is that because $\kappa_{w}\gg \kappa_{m}$, an ion's electric field lines are concentrated inside the channel over a length proportional to $l_1 \sqrt{\kappa_w/\kappa_{m}}$, where $l_1$ is the shortest dimension of the channel \cite{Teber}.  The specific form $U_{S}$ takes depends on the nature of the channel. For a planar channel the electrostatic potential varies as $U(r) \sim \ln r$ for length scales $l_1 \sqrt{\kappa_w/\kappa_{m}} > r > l_1$, while for a linear channel $U(r) \sim r$.  For channels which are relatively short and narrow, the larger dimension of the channel $L \simeq l_1 \sqrt{\kappa_w/\kappa_{m}} \gg l_1$; this implies that the self-energies scale as $\ln (L/l_1) / l_1$ and $L/l_1^2$ in planar ($l_1 \times L \times L$) and one-dimensional ($l_1 \times l_1 \times L$) geometries respectively. For example, for a water-filled  channel of dimensions $1 nm \times 1 nm \times 5nm$, $U_S$ is about $7$ $k T$ at $T=300$K where $k$ is the Boltzmann constant \cite{Shklovskii_PRL2005}.

In addition to electrostatic interactions, one might inquire as to the importance of hydrodynamics interactions. It is easy to see that hydrodynamic interactions are important only for systems which are much larger than some characteristic scale $R_*$. The length scale $R_*$ may be estimated by comparing the electrostatic and the hydrodynamic forces between two ions separated by a distance $r$. The electrostatic interaction (in three dimensions) is $f_{E}=\frac{e^{2}}{4 \pi \epsilon_0 k_w r^2}$ while the hydrodynamic force is $f_{H}=\gamma u_d r_0/r$, $\gamma$ is the viscous drag coefficient, $\epsilon_0$ is the dielectric constant of the vacuum, $r_0$ is the radius of the ion, $e$ is the charge of the ion and $u_d$ is the ion drift velocity. Taking $u_d= (E/\gamma)$, where $E$ is the electric field acting on the ions in the channel, we obtain $R_*=e/(4 \pi \epsilon_0 \kappa_w r_0 E)$. For the experimental conditions of \cite{Shklovskii_PRL2005}, $R_* \approx 10 nm$ which is larger than the channel scale (the same result holds in two dimensions). In this paper our interest is in this regime, consequently we will ignore hydrodynamic interactions. 

Non-monotonic behavior in charged channels was previously studied theoretically using a single vacancy model \cite{Schumaker-MacKinnon}, under the assumption that the channel was strictly one-dimensional.  A more recent study considered the ion current in a channel threaded with charged DNA, where the available space for ion motion was assumed to be effectively two-dimensional. In this case, the non-monotonic behavior was attributed to the two-dimensional specifics of the channel and the self-energy of the ions, and to a boundary layer effect at the edges of the channel \cite{Bonthuis}. 

\begin{figure}
\begin{center}
\leavevmode
\includegraphics[width=5.0cm,angle=0]{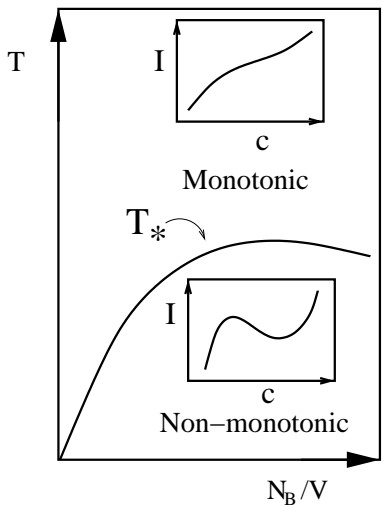}
\caption{Schematic phase diagram of ion transport in a channel of confined geometry at temperature $T$ and volume $V$ with $N_B$ immobile background charges inside the channel. The thick line denotes the crossover temperature $T_*$. The two insets are plots of ion current $I$ versus external ion concentration $c$.}
\label{phase-diagram}
\end{center}
\end{figure}

In this paper we present a many-particle statistical model of interacting ions, and argue that, in the presence of fixed background charges inside the channel, the large self-energy of an individual ion is sufficient to give rise to a non-monotonic ion current $I$ as a function of external ion concentration $c$.  Our main result is that, irrespective of the effective channel dimension, there is a crossover temperature $T_* \simeq U_S / k$, below which the ionic current may exhibit non-monotonicity.  However, above $T_*$, the current is a monotonically increasing function of $c$.  Consequently, non-monotonic behavior can be observed only when $U_S$ is large enough for $T_*$ to be above the freezing temperature of water. For example, when $U(r) \sim 1/r$, as in large three-dimensional cavities, $T_*$ is much below the freezing temperature of water but when $U(r) \sim \ln r$, the $I$ {\it vs} $c$ curve may have a minimum even at room temperature. In any case, for very high (or very low) density of background charges, $I$ increases monotonically with $c$ as is naively expected. This is summarized in Fig.\ref{phase-diagram}.

The above results are the consequence of two main competing mechanisms for ion transport: (1) {\it Hopping current} $I_{h}$: At low temperature, the fixed background charges are screened by counter-ions \cite{Rabin}, which thus reside in close proximity to the background charges - one may think of the counter-ions as sitting `on the sites' of the background charges.  However, if one of the background charges is not screened (a `hole'), the screening counter-ion of an adjacent background charge can hop to it.  $I_{h}$ is approximately proportional to $\rho_h(\rho_0-\rho_h)$, where $\rho_h$ is the density of holes, and $\rho_0$ is some constant. Since $\rho_{h}$ decreases with increasing $c$, the ion current first increases, attains a maximum (at $\rho_{h}=\rho_{0}$) and then decreases. (2) {\it Bulk current} $I_{b}$: Ions that are not strongly attached to any counterions will move more or less freely inside the channel, and, biased by the electric field, will contribute to the total current. $I_b$ is a monotonically increasing function of $c$. The total ion current $I$ is sum of the hopping current $I_h$ and the bulk current $I_b$, $I=I_h+I_b$. 

This intuitive picture for $I_h$ is supported by a simple model for driven diffusion, the {\it partially asymmetric simple exclusion process} (PASEP) \cite{Derrida}. The PASEP considers a one-dimensional lattice of sites, each of which may be either empty or occupied by a single particle.  Particles may enter or leave the system at its ends,
and a particle may hop to an adjacent site provided it is unoccupied.  The parameters of the model are the rate of influx ($\alpha$, $\gamma$) and outflux ($\beta$, $\delta$) of particles at the left and right ends, respectively, and the hopping rates between sites: $q<1$ and $1$, to the left and right, respectively (where the applied electric field may be thought of as the cause of asymmetry of the hopping rates).  In the ion channel, a fixed charge screened by a counter-ion maps to an occupied site in the PASEP model, and an unscreened fixed charge maps to an unoccupied site in the PASEP. 

The phase diagram of the PASEP model has been fully elucidated (see for example, \cite{Derrida,Uchiyama}).  If the incoming rates $\alpha$ and $\gamma$ are taken to be proportional to the outside concentration $c$, the behavior of current $I_{h}$ can be immediately obtained using these results.  It follows from \cite{Uchiyama} that $I_{h} \sim c$ for small $c$, and $I_{h} \sim 1/c$ for large $c$. At intermediate $c$, the current attains a maximum or a plateau.  In the PASEP model, the various rates are taken to be constant, but, in reality, rates will depend on specific configurations of the system. Clearly the PASEP model cannot capture the appearance of the minimum in the $I$ {\it vs} $c$ curve.

To understand this minimum we will consider a statistical mechanical model of interacting ions in an ion channel where the channel is  in contact with a reservoir of a fixed chemical potential $\mu$ and temperature $T$.  For simplicity, we will consider a discrete model, where the positions of ions lie on a lattice.  The kinetic energy of the ions is neglected, since ion motion in a fluid is overdamped. We assume that the electrostatic potential $U(\vec{r})$ of a unit positive charge at position $\vec{r}$ inside the channel decays rapidly outside the channel. The Hamiltonian for a system of $N$ interacting charges of hardcore radius $r_0$ is
\begin{equation}
H = \frac{1}{2} \sum_{i\ne j} q_i q_j U(r_{ij}) + \frac{1}{2} U_0 N - \mu N
\label{H1}
\end{equation}
where $q_i=\pm 1$ is the charge of $i$-th ion, $U(r_{ij})$ is the interaction potential of ions $i$ and $j$, whose separation is $r_{ij}$, $N$ is the total number of ions and we denote $U_0 \equiv U(r_{ij}=r_0)$; the self energy of an ion is given by $U_S=U_0/2$. The definition of the Hamiltonian absorbs the chemical potential $\mu < 0$, for simplicity assumed to be the same for both positive and negative charges, which is related to the fugacity $z$ by $z=\exp(\mu/kT)$. Note that inserting a bound neutral pair (one +, one - charge) costs an energy $-2\mu$: because of cancellation there is no contribution from the first two terms in Eq. \ref{H1}.  The fugacity $z$ controls the density of ions inside the channel.  

For small external electric fields, it is reasonable to assume that local thermal equilibrium is maintained.  We  thus include a constant external electric field $E\hat{x}$ along the channel axis. 
Using $\sum_{i\ne j} q_i q_j = [(\sum_i q_i)^2-\sum_i q^2_i]$, Eq. \ref{H1} may be rewritten 
\bea
\nonumber
H = \frac{1}{2} \sum_{i\ne j} q_i q_j [U(r_{ij}) - U_0]  -  E \sum_i q_i x_i
\\
- (N_+-N_- - N_B)^2 k T \ln (z_b) - N k T \ln (z)   
\label{H_immobile_ions}
\eea
where we have explicitly indicated the $N_B$ fixed negative background charges, and where the sum is over all pairs of ions except those where both are background charges. Here $N_+$ and $N_-$ are the total number of positive and negative mobile ions respectively, $N=(N_+ + N_-)$, $x_i$ the x coordinate of $i$-th mobile ion, and $z_b\equiv \exp [ -U_S/kT ]$. For small $z_b$, charge fluctuations in a finite channel are small, and $(N_+-N_- - N_B)\simeq 0$ \cite{Rabin}. 

We begin by considering the system at zero electric field; the charge distribution is then governed by the partition function ${\cal{Z}} = \sum ({1}/{N_+!N_-!}) \exp(-H_{0}/T)$, where $H_{0}$  is the Hamiltonian of Eq. \ref{H_immobile_ions} with $E=0$ and the sum is over all configurations.  For small electric fields, the charge distribution will be essentially unchanged; we will use this to calculate the ionic current.  

Let us consider the fugacities $z \sim z_b \ll 1$. Here we expand ${\cal{Z}}$ in powers of $z$ and $z_b$. Collecting leading order terms, we obtain
\bea
\nonumber {\cal{Z}} \simeq N_B z^{N_B-1} z_b + z^{N_B} + {\cal O} (z^{N_B-2} z_b^4)  
\\
+ {\cal O} (V z^{N_B+1} z_b)  + {\cal O} (V z^{N_B+2})
\label{Pfnseries1} 
\eea
where $V$ is the channel volume measured in units of ionic volume.  Numbering the terms on the right hand side of Eq. \ref{Pfnseries1}, we may interpret them as follows: (1) one unscreened immobile charge (one hole), (2) all immobile charges are screened (no hole), (3) two unscreened immobile charges (two holes), (4) one excess positive or negative charge (apart from the screened backbone charges) and (5) one excess {\it bound} pair of positive and negative charges. In terms (4) and (5), the factor $V$ accounts for the possible placements of the extra charges. Eq. \ref{Pfnseries1} can be well-approximated by the first two terms alone for $z_b \sim z \ll z_*$, where $z_{*} \simeq min\{V^{-1/2}, (z_b N_B/V)^{1/3}\}$. Now in this fugacity range, the probability $P_h$ that there is exactly one hole can be written as $P_h \approx {N_B z_b}/{(N_B z_b + z)}$ using Eq. \ref{Pfnseries1}.
 
Consider first the behavior at low fugacity, $z < z_b$, which we will call {\it Region I}. By examining Eq. \ref{Pfnseries1} one can see that as long as $z$ is not extremely small, the dominating configuration has one uncompensated background charge. The current flows by positive charges hopping from one background charge to another so that the hole moves from one end of the system to the other. The probability $P_h$ of having one hole in the system depends weakly on $z$ in this regime. For the current to flow the hole must recombine with a charge from outside the pore, and this occurs with a recombination rate proportional to $z/z_b$. The current $I$ is proportional to $P_h$ time the recombination rate, giving $I \approx I_h \sim z$.

Now consider intermediate fugacities $z_b < z \ll z_*$. In this regime, which we term {\it Region II}, the probability $P_h$ of having one hole goes as $1/z$. Since for $z > z_b$ the recombination rate can be approximately taken as $1$, the hole current, which in this regime is proportional to the hole density, is therefore given by
\begin{equation}
I_h \approx \sigma \times \frac{z_b}{(N_B z_b + z)}
\end{equation} 
where $\sigma$ is a constant related to the jump-rate of a hole from one site to another. In this fugacity range, there are no free bulk charges, so the total current $I \approx I_h \sim 1/z$, decreasing with increasing fugacity. 

{\it Region III} is the large fugacity limit $z \gg z_*$, where extra charges enter the system, although the background charges are already fully compensated. In this regime the current is clearly expected to increase with increasing fugacity.

As a function of increasing $z$, we have the following: The current rises linearly in Region I, falls as $1/z$ in Region II, and rises again in Region III.  Thus, it is the passage from Region I to II that determines the non-monotonic behavior.  However, Region II may be unobtainable - this happens when $z_* < z_b$.  In this case, Region I crosses smoothly over to Region III, and the ion current increases monotonically with $z$ over its entire range. In other words, Region II is present only if $T < T_*$, where $T_* = 2U_S/k\ln (V/N_B)$.  Note that $T_*$ increases with the number of bound charges $N_B$. The above picture breaks down when the density of background charges is so high that 
ions can move freely (without hopping) from one background charge to another. This occurs when the typical distance between background charges is smaller than the screening length. Under such conditions we expect the current to increase monotonically with fugacity.

To support these simple arguments, we have performed Monte-Carlo simulations.  For computational convenience, ions are only allowed to move in discrete steps on a square lattice.  Ions can enter and leave the system only from two opposite surfaces, representing the pores of the channel. A site may accommodate at most one ion. We denote by $\Delta H$  the energy difference between configurations after and before a possible Monte-Carlo move, with $H$ defined in Eq. \ref {H_immobile_ions}. The simulation is carried out in the following way:  At each time step, a lattice site is randomly chosen.  If it is an empty boundary site, a positive (negative) charge is created with probability $min\{\frac{1}{2}, \frac{1}{2} e^{-\frac{\Delta H}{kT} } \}$.  If the site is occupied, the charge is destroyed with probability $min\{1, e^{-\frac{\Delta H}{kT} } \}$.  If the site is in the interior of the lattice and is occupied, its charge is moved to a randomly chosen unoccupied neighboring site with probability $min\{\frac{1}{4}, \frac{1}{4} e^{-\frac{\Delta H}{kT} } \}$. For $E=0$, the system eventually comes to  equilibrium, while for $E \ne 0$, the system settles into a non-equilibrium steady state with a net ion current across the channel in the $x$-direction.  

Motivated by the experiment of Ref. \cite{Bonthuis} which is effectively two-dimensional, we performed a simulation on an $L \times L$ lattice using the above protocol, with the interaction potential taken to be $U(r) = ({2 e^2}/{\kappa_w r_0}) \ln (L/r)$, where $e$ is the electron charge, and  $r_0 \simeq 0.35$ nm \cite{Bonthuis}.  An immobile linear array of equally spaced unit negative charges is placed on a line parallel to the $x$-axis in the middle of the channel, at $y=L/2$, to mimic the presence of charged ss-DNA in the experiment.  One should note that when $L$ is large and $N_B=0$, Eq. \ref{H_immobile_ions} is the 2D Coulomb gas Hamiltonian \cite{Minhagen}.  

\begin{figure}
\begin{center}
\leavevmode
\includegraphics[width=7.5cm,angle=0]{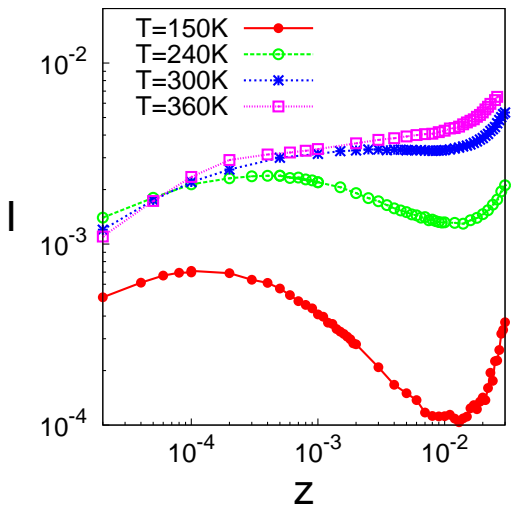}
\caption{Ion current $I$ (in arbitrary unit) across a two-dimensional channel versus fugacity $z$ is plotted for different temperatures with $L=30 \times d_{K^+}$ where diameter $d_{K^+}$ of $K^+$ ion being $0.26$ nm, an electric field $E=5.4$ meV/nm along x-axis, $N_B=6$ negative immobile background charges.}
\label{crossover_ND6}
\end{center}
\end{figure}

The results of the simulations are presented in Fig. \ref{crossover_ND6} where we plot the total ion current versus fugacity for different temperatures. As expected from the arguments presented above, the numerical results are qualitatively different in two different temperature regimes. For $T < T_* \approx 300 K$, the ion current $I$ first increases with $z$ for small $z$, then reaches a maximum and subsequently decreases.  Increasing $z$ further, $I$ reaches a minimum and then starts increasing with $z$. For $T > T_*$ the current $I$ is a monotonically increasing function of fugacity $z$. The numerical value of $T_*$ given above is somewhat smaller than that given by $T_* = 2 U_S/(k \ln (V/N_B)) = 816 K$. This is about what could be expected from such a simple argument.

\begin{figure}
\begin{center}
\leavevmode
\includegraphics[width=8.0cm,angle=0]{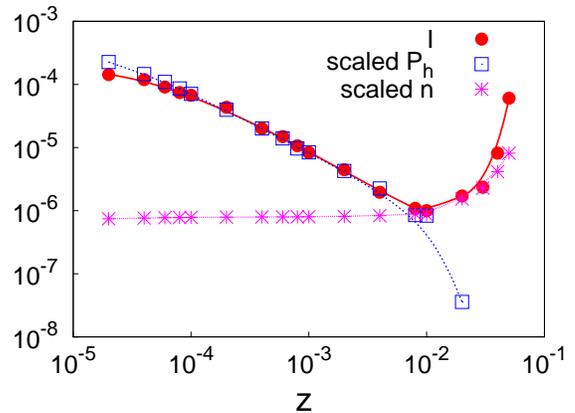}
\caption{Current $I$ (in arbitrary unit), scaled probability $P_h$ of exactly one hole (scaling factor $0.00055$) and scaled total number of ions/site $n$ (scaling factor 0.00012) is plotted versus fugacity $z$ with an electric field $E=1.1$ meV/nm, temperature $T=150K$, $L=30 \times d_{K}$ where $d_{K}=0.26$nm is the diameter of a $K^+$ ion, $z_b=1.23 \times 10^{-6}$ and $N_B=6$ negative immobile background charges. }
\label{fit_prob_one_hole}
\end{center}
\end{figure}

In Fig. \ref{fit_prob_one_hole}, we plot numerical results for the ion current $I$, the probability $P_h$ that the system has exactly one hole, and the average total number of ions per site $n$ as a function of the fugacity where both $P_h$ and $n$ are scaled suitably to relate to $I$ for $T < T_*$. As can be seen at low fugacities ($z \leq 0.01$), single hole hopping is responsible for the ionic current. For larger fugacities, the number of free bulk charges increases, and the current, almost entirely due to flowing ions in the bulk, rises.  For large self energies, as in the simulation, inserting a pair of positive and negative ions into the channel is much easier than inserting a single charge, so the number of unbound charges inside the system increases as $z^2$ for $z_* < z \ll 1$.  This is seen in Fig. \ref{fit_prob_one_hole}, which shows a concomitant rise in ion current. 

Finally, it is worth noting the influence of the effective dimension of the system, which manifests itself in the functional form of the Coulomb interaction.  In Monte-Carlo simulations of the same geometry (a $30 \times 30$ lattice with a linear array of negative immobile charges in the middle), but employing a Coulomb interaction $U(r)=1/r$, we found that the minimum in the current-fugacity plot can, in principle, also occur, but only at very low temperatures, of order $T \sim 36$ K which is clearly experimentally irrelevant.

DL and YK acknowledges support from the Israel Science Foundation under grants 1574/08 and 1183/06. PP acknowledges support from the Russell Berrie Nanotechnology Institute at the Technion.

\end{document}